# LASER POINTER PROHIBITION – IMPROVING SAFETY OR DRIVING MISCLASSIFICATION
Paper #101

Trevor A Wheatley

School of Engineering and Information Technology, UNSW Canberra,
Campbell, ACT, 2612, Australia

## Abstract

It is well known that since 2008 Australia has had some of the world's most restrictive laws regarding the possession and importation of "laser pointers" with powers exceeding 1 mW. Now four years on Australia is used as a test case and question whether this has actually improved safety for those wishing to purchase these devices or if it has impacted on the availability of prohibited devices. Results from the analysis of over 40 laser pointers legitimately purchased in Australia from local and International suppliers are presented. Specifically lasers that are readily available to everyday consumers through the simple on-line search "laser pointer 1mw" are targeted. The parameters investigated are quoted power versus measured power, correct representation in advertising and adherence to laser standards as related to specified use and purchase price. The analysis indicates that the suppliers in this market have learnt how to bypass the prohibition and the impact on general safety in these cases is detrimental.

## Introduction

Laser pointer access and misuse is arguably one of the most emotive and controversial issues currently facing regulators, laser safety professionals and consumers. Many countries have ongoing problems with laser pointers being directed at aircraft, motor vehicles and members of law enforcement. Having had many conversations on the issue with various members of the laser community, it would seem that there is not a common opinion on how to remedy this situation. In fact there isn't even common definition of the term "laser pointer". Commonly the definition is related to the form factor and the power rating, but this is not always the case and not the case here. To avoid confusion with respect to the terminology, it is prudent to clarify the use of the term "laser pointer". For the purpose of this paper the term laser pointer refers to any hand-held battery operated laser device that produces a beam of any power that may be used for aiming, targeting or pointing.

The approach taken by Australia in 2008 to mitigate this misuse was a hot topic for discussion in various forums and it is still a polarising issue within the laser community. Whether this approach was justified, fair, sensible, enforceable or correct is often the topic of these discussions, but opinions aside it was a proactive step to attempt to counter the misuse. The question now, and the motivation for this paper, is whether it has achieved its intended purpose?

## Background information

Misuse of laser pointers particularly with respect to aircraft has been a concern for Australian regulators for some time, culminating in substantial law changes that came into effect in 2008 [1,2]. In 2006 the Civil Aviation Safety Authority (CASA) published an internal article in relation to laser pointers and aircraft. The introduction of laser pointer import restrictions was raised Federally by the Premier of Victoria who wrote to the Prime Minister on the matter in 2007 [1]. However, it was the high profile media attention given to a series of laser incidents in March 2008 at Kingsford Smith Airport in Sydney that prompted more immediate action. This action was taken at both State and Federal levels of Government and resulted in the current legislation regarding import and possession of laser pointers.

In order appreciate the legislative approach taken; a basic idea of the relevant Australian Government framework is useful. Australia has both Federal (Commonwealth) and State Governments [3], both of which in this context (weapons control) have different responsibilities. The Federal Government is concerned with controls and restrictions associated with import and export, implemented through Australian Customs and Border Protection Services (referred to as Customs henceforth). The State Governments on the other hand are responsible for regulations relating to possession and use; this usually administered by the police through licensing arrangements within their jurisdiction. As there are a number of independent State Governments the individual details of the resulting legislation vary, but for the most part the prohibition or control applies to "a hand-held battery operated laser pointing device, with a power exceeding 1 mW" [2, 4 – 6]. It should be noted that these devices are not banned outright in any State or Territory; they are subject to legislative control and usually require some sort of permit to possess. Thus most States have exemptions that allow legitimate users, for example astronomers, to obtain these devices. These exemptions would appear to be routine for reasonable use of laser pointers with an output power below 20 mW, but whilst exemptions for higher output power are available they are less routine. The two tier Government approach is reflected in the import approval process which typically involves approval from the both State and Federal Government bodies [7].

The sale of laser pointers like all consumers product sold in Australia is subject to consumer law. As is the case in most countries, businesses that sell unsafe or defective products in Australia may be subject to prosecution. One way for manufacturers to mitigate this risk is by compliance with product safety standards, both mandatory and voluntary standards are used in Australia. Manufacturers can be called on by regulators to produce evidence of testing and standards compliance if a product is investigated. In addition to general consumer law requirements the sale of laser pointers exceeding class 2, was already restricted in some Australia States prior to the 2008 law changes [8]. These restrictions were brought about as a result of accidents [8] or perceived risk that class 3 [9] and 4 lasers were potentially hazardous and not recommended as consumer products. Thus the new laws have often been perceived as an over regulation of these devices and an overreaction due to the media coverage (see any number of laser pointer forums). However, Federally one could argue that without import restrictions, a constraint on Australian businesses selling laser pointers would do little to discourage International entities selling these devices to Australians. It is acknowledged that more could be said here the on merit of these prohibition laws, for further discussion and opinions see the aforementioned laser pointer forums. The intent of this paper to look not at merit of these prohibition laws but whether they have made obtaining a laser pointer in Australia more difficult and what is the impact on laser safety.

### Research scenario

It was mid 2012, the "1 mW 405 nm laser pointer" purchased on EBay from an Australian seller had arrived. Whilst assessing if the device met with expectations its power was measured and found to exceed the 1 mW threshold substantially. The laser was also not correctly classified in accordance with Australian standards nor any standard for that matter and was mislabelled. Thus posing questions such as:

- Is this an anomaly?
- Did the seller deliberately misclassify and incorrectly advertise it?
- Are the strict State and Federal laws actually limiting access to these devices?
- Is there a laser safety concern for the less sceptical consumer without measurement equipment?

This required further investigation and hence the motivation for this work. The scenario considered in this paper is that of the average consumer wishing to purchase a low cost laser pointer for any number of legitimate reasons. It assumes that the average Australian consumer is internet savvy and aware of the legislation relating to laser pointers insofar as the 1 mW limit is concerned. Thus an average consumer may go to Google and enter "laser pointer 1mW"; this was the basis for the laser pointers purchased for this research. Upon entering the above search parameter into www.google.com.au, the first (non-paid) result was followed. This link was to www.ebay.com.au, with "laser pointer 1mw" entered into the search engine. The results indicated 105 hits for this search, this was narrowed down to 83 "buy it now" items and further reduced by not considering identical listings from the same seller (i.e. same form factor and colour). The remaining 44 (17 Australia, 13 Hong Kong, 11 China, two UK and one USA [10]) laser pointers of three colours (20 red, 18 green and six violet) were purchased for analysis.

### Results

The first and simplest question was, of the lasers purchased how many were delivered? Of the 44 lasers purchased 40 were delivered without incident, three were intercepted by Customs and one was not received (noting there no response from the seller to enquiries). Of the three laser pointers intercepted and assessed by Customs, two were measured to exceed 1 mW (results were provided by Customs) and were retained for disposal. The remaining laser pointer was measured to have an output of less than 1 mW and was released for delivery. So of the original purchase of 44 lasers, 41 were actually received.

All lasers where advertised as 1 mW (or less), however upon arrival the stated power on the label was ≤ 1 mW or ≤ 5 mW 68% (40% & 28% respectively) of the time, a significant number (30%) where not labelled at all and one product stated < 10 mW. It would appear that the International suppliers are targeting a 5 mW restriction for laser pointers but advertising as 1 mW for the Australian market on the ".au" domain. On the other hand the Australian based sellers were labelling to feign compliance with the 1 mW prohibition limit with 88% labelled as ≤ 1 mW. However, the labelled class was often not consistent with stated power and in all cases the Australian (or IEC) Standard. In a significant amount of cases lasers carried labels indicating class III, which does not reflect a current class of laser product. Based on these inconsistencies one would be justified in questioning the quality of information provided to purchaser, motivating the need for measurement. On a side note an unexpectedly high number of these sellers advertised their products as "lazers", perhaps a consumer's first warning.

Probably the most damning piece of evidence from a laser safety perspective came when the measurements were taken. The purpose of the measurements in this case was not to achieve fine precision but to confirm to that the claimed power levels were not overtly exceeded, so in-house equipment was deemed suitable. Nevertheless, a formal approach using a Newport 1918-c with a calibrated (Cal date 01 Aug 2012, cal expire 01 Aug 2013) Newport 918D-SL-OD3 detector head was taken. Additional validation measurements were taken with Thorlabs PM100D meter using an s310C detector head (cal date not specified). A series of 5 independent measurements were taken with each meter and the average power and variance taken as the power measurement and the uncertainty respectively. It should be noted the variances associated with the green lasers were typically higher than the red and violet lasers, most likely due to the temperature dependence of the second harmonic generated outputs. Also for the green lasers filters were used to allow separation of the fundamental, pump and second harmonic, the sum of the measurements was compared with the claimed output power. Loss through the filters was approximately 3% and was accounted for at the fundamental and second harmonic wavelengths. The correction at the pump wavelength was considered negligible due to small emissions. A photograph of the measurement setup is shown at figure 1.

A bar graph showing the most significant measurement results is at figure 2, with first 22 lasers shown in the upper section and the second 22 in the lower, the lasers are numbered according to their purchase order. Figure 2 shows that there is little correlation between the advertised (1 mW) and the measured laser outputs, the green, red and violet bars show the average power in mW by colour. This data also shows (see blue bars scale to be read in dB relative to the MPE) that in spite of being advertised and sold as laser pointers that do not exceed the MPE (1 mW for small source for visible lasers), disturbingly all but two (see negative blue bars for lasers 13 and 34) of these lasers exceed the MPE. A significant number of these lasers exceed the MPE by upwards of 10 times (see blue bars exceeding 10 dB).

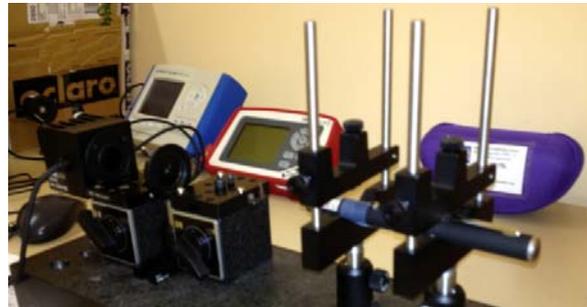

Figure 1 The measurement setup for this work, magnetic optics mounts were used to stabilise the laser pointers and the meters. The variation in form factor of the devices meant that flexible supports were required.

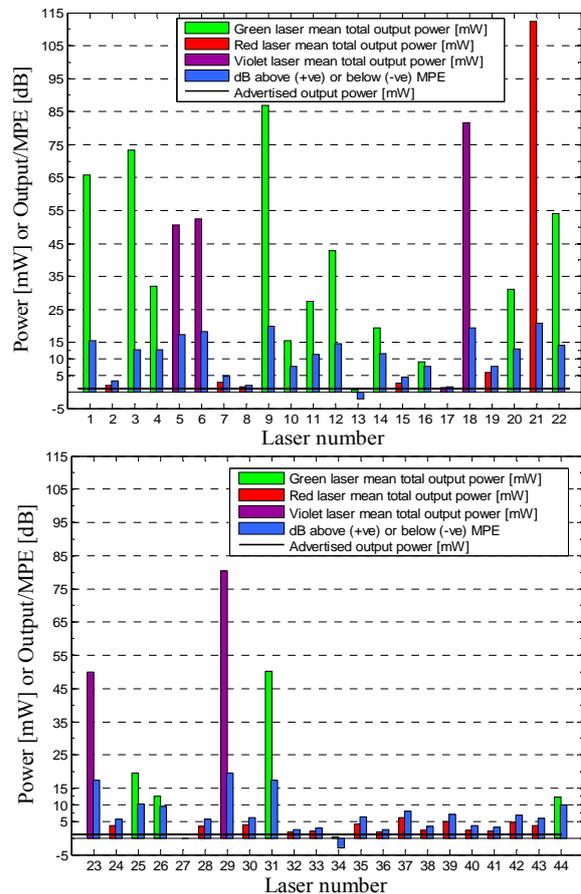

Figure 2 Mean measured laser output power by colour (green, red and violet bars) and ratio of the additive maximum power to MPE power (blue bars) shown as dB above or below the MPE as a function of laser purchase number. The black line at 1 mW is provides a reference to the expected maximum power output.

The data seem to indicate that the green and violet lasers represent the most significant hazard; however, the laser with the highest power output was a red laser (number 21) measured at 112 ± 8 mW. It is also noted that all of the red lasers from laser number 24 onwards were incorporated in other devices such as wireless keyboards, PowerPoint remotes or LED torches (see figure 3), most of which use button batteries and in general do not appear to have the capacity of the stand alone laser pointers. This seems to indicate that these red lasers are lower risk, but it should be still noted that whilst these lasers did not represent as significant a hazard as the single purpose pointers, none of them performed as advertised and all exceeded the 1 mW limit, indicated by the black line on figure 2.

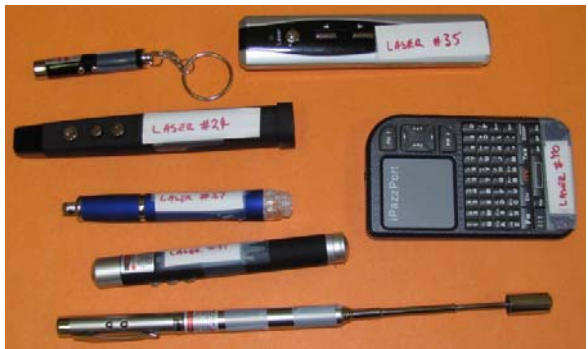

Figure 3 A example of were incorporated in other devices that were captured by the search parameter for this research.

The worst offending lasers in this study with respect to safety are those with the black and silver, two AAA battery form factors shown in figure 4. This packaging style is not restricted to any particular laser colour but it was noted that lasers in this packaging were invariably non-compliant and usually higher risk. A significant percentage (78%) of lasers in this form factor exceed the MPE by between 5 and 100 plus times. In fact only one such laser did not exceed the MPE in this case study.

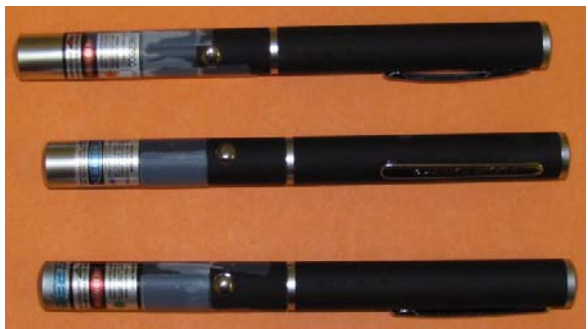

Figure 4 Three laser pointers purchased for this research that are indicative of the mostly likely form factor for the high risk lasers.

Figure 2 paints the picture with respect to the Australian legislation, with all but 2 exceeding the 1 mW limit. However, the typical perception is that the risk for accidental exposure does not become too significant until the output exceeds 5 times the MPE [11]. Another significant contributor to risk of injury in this context is purchase price, as the cheaper the product the more attainable it is for the general public and children. The assumption here is that the average person probably doesn't wish to spend significantly more than about 20 Australian dollars (AUD) on a novelty product or a cat toy. Figure 5 shows the purchase price by colour and laser purchase number. It is apparent that the majority of these lasers were less than 20 AUD and so they would seem very accessible. The plot shows two lasers (34 and 36) that were significantly more expensive than the others. It would seem that the extra expense for laser 34 was for engineering and testing, as it was the only one of the entire sample that performed as claimed and was compliant with laser safety standards. Whilst the expense seems to be justified for laser 34, quite the opposite was the case for laser 36 which was the poorest quality out of the sample set, and the extra expense cannot be reconciled. However, it would appear as though there is little correlation between colour and purchase price, even additional features such as wireless technology didn't make the cost prohibitive.

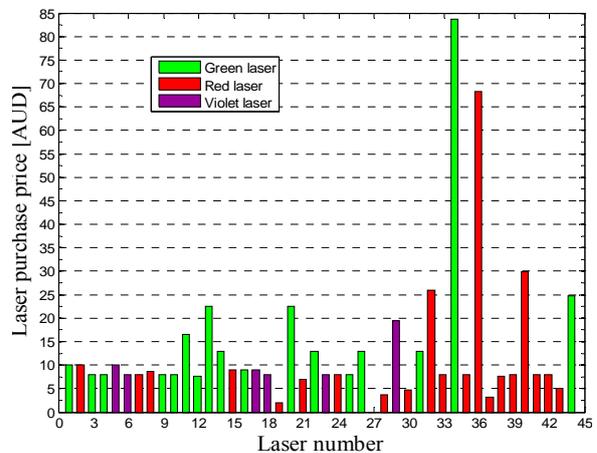

Figure 5 Laser purchase price in Australian dollars including shipping showing colour as function of purchase number.

It can be seen in figure 5 that most of these lasers are not out of reach to the average person or child by way of price. Presenting this from another perspective figure 6 brakes the purchase price into 5 AUD bins and shows the number of lasers (green bars) whose purchase price fell within a given range. Additionally if we consider lasers that exceed the MPE by greater than 5 times as moderate to high risk (ie class 3B power levels); it can be seen that this is about 53% (23 of 43) of the purchased "safe" lasers. Further if we assume that lasers costing less than 20 AUD are accessible to children and hence higher risk, we find that approximately 60% of these lasers are of class 3B power levels. Based on the fact that 92% of the lasers purchased for less than 20 AUD were delivered, there would appear to be a greater than 50% chance that someone attempting to buy a low cost "safe" laser pointer would inadvertently get a hazardous laser. This is not to mention that 100% of the aforementioned sub 20 AUD lasers would represent prohibited weapons in most Australian States. Importantly if we consider access by children, then exceeding class 1 is an unacceptable risk so even the 1 mW prohibited weapon threshold is probably too high [12].

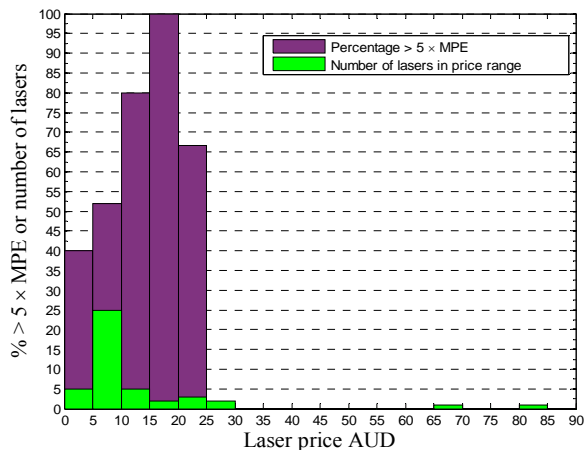

Figure 6 Percentage of lasers in a price range that exceed the MPE by 5 times (purple) and the number of laser samples with that price range (green) as a function of five Australian dollar increments.

## Discussion

The results indicate that if a consumer in Australia attempted to purchase a low cost compliant laser pointer in this manner, they would most likely not get one. In fact if they restricted their purchase to one without other features such as the ones shown in figure 4, they would likely receive a potentially hazardous device. The stated intent of the prohibition is "to limit the availability of certain laser pointers within the Australian community and play a part in reducing the number of incidents of misuse" [1]. The statistics provided in [1], state that there was 648 laser incidents involving aircraft in 2007/08 financial year (July-June), this deceased in the first year after the prohibition but began increasing again in the following year reaching 828 in 2010/11. This would seem to indicate that availability has not been significantly impacted. [1] also states that there were 12,457 lasers intercepted at the border in the year after prohibition was introduced and that this number has increased by about 12,000 per year for the 3 years shown. If the percentage intercepted in the case study is not an anomaly, the actual number of prohibited laser pointers entering the country may have been high as half a million last year. So whilst it is easy to suggest that without these controls the numbers would be much higher, it is hard to say that this approach has been a resounding success.

The analysis of the lasers in this case study indicates that purchasing in this way will likely result in a more hazardous than expected laser and that the suppliers are aware of their market. The former has been discussed in the results section; however the latter warrants further explanation. It is believed that the fact that all these lasers where claimed to be either less than or equal to 1 mW on the ".com.au" version of the website is not a coincidence. It is certainly very likely that these suppliers (particularly the Australian based ones) are aware of the import and possession laws and so advertise these products to give the impression that they are legitimate products. Reinforcing the argument that the suppliers are deliberately misrepresenting these products to get around the legal controls is the manner in which these products are packed for shipping. Of the 43 laser pointers that were either received or intercepted (photographs of the packaging were supplied by Customs) only one was clearly labelled as a laser pointer. It is noted that the one package labelled as a laser pointer was intercepted by Customs but as it was also the only compliant product, it was released for delivery. This incorrect marking of packaging by suppliers when coupled with the vast number of legitimate small electronic packages entering the country must make the task of intercepting a large percentage of laser pointers is very difficult.

The intent of this research was not to make the case against prohibition as more than likely it has discouraged a number of consumers from purchasing laser pointers. Moreover, it is fair to say that it has raised the profile of the hazards associated with high power laser pointers and their misuse. Unfortunately this publicity may well have served as inspiration for some to perpetrate similar acts of misuse. It is also fair to say that those who are of the mindset to point lasers at aircraft are probably not deterred by the fact that their laser pointer is prohibited. One mechanism by which this legislation may have been effective is by restricting the access to such laser pointers. It is in this area where it is seems that the prohibition and import restrictions have not appeared to have made a significant difference. The answer to the question of how to remedy this, if at all, is not obvious. However, what seems to be apparent from this case study is that a single pronged approached might be insufficient. A coordinated approach, additionally involving the manufacturers of these devices and the websites through which purchases are possible may provide a solution. An example of website restrictions making access more difficult was witnessed during this case study [10]. The purchase of this laser pointer, which was being sold by a US supplier, was blocked by the website and the transaction not able to be completed. It is recognised that the work-around for this was relatively easily achieved as there was no coordination of effort, which in practice would be difficult maybe impossible to achieve.

## Conclusions and Summary

For all except the one laser received from a US supplier there appeared to be little correlation between the advertised/labelled power and the actual power of the device. In fact the results indicate that little to no effort is actually taken to test these devices against any laser safety standards and the marketing is aimed at the target market.

The results indicate that high price (> 70 AUD) is potentially an indicator of likely compliance but this warrants further investigation. However, for low cost lasers (< 20 AUD) there is no evidence to suggest that there is any correlation between output power and price.

There seems to be some relationship between device purpose and output power, with the typical laser pointer with other functions, e.g. wireless capability, having lower output powers than single purpose laser pointers. However, it is noted that this could be by virtue of the type of batteries and laser diodes used, as all had red laser diodes and most used button batteries. Noting that with respect to the question of accurate representation and standards compliance they still failed to meet expectations.

In relation to colour it was found that green and violet lasers were more likely to significantly exceed safe power levels. However, this was more likely a function of the types of products that use green and violet light rather than the colour itself, as red lasers in the same form factor also significantly exceeded the safe limits.

In conclusion it is thus postulated that the prohibition laws may have detrimentally affected laser pointer safety within Australia without overtly impacting availability. This statement is based on the observation that these lasers are being marketed as legal, safe and less than 1 mW but are generally anything but legal, safe or less than 1 mW. Or viewed from a laser safety perspective: *the one thing more hazardous than a correctly labelled high power laser pointer is a high power laser pointer labelled as safe.*

**Acknowledgements**

The author wishes to acknowledge the resources provided by the School of Engineering and Information Technology, UNSW Canberra and Australian Customs and Border Protection Services for the provision of measurement information and packaging photos for seized lasers.

**Meet the Author**

Trevor Wheatley is an Electrical Engineer with tenure as a lecturer within the School of Engineering and Information Technology UNSW Canberra, where he researches in experimental quantum parameter estimation. Trevor chairs the Standards Australia SF019 Committee on laser safety and is the head of the Australian delegation on IEC TC76. He is the presiding member of UNSW Canberra laser safety committee. Trevor researches, teaches and consults in laser safety in both Australia and New Zealand.